# Role of phase fluctuation in dynamic competition between charge order and superconductivity in cuprates

Mingu Kang[1,2,10], Pavel E. Dolgirev[3,10], Chao C. Zhang[4], Hoyoung Jang[5,6], Byungjune Lee[2,7], Minseok Kim[5], Sang-Youn Park[5], Ronny Sutarto[8], Eugene Demler[9], Jae-Hoon Park[2,6,7], John Y. T. Wei[4] & Riccardo Comin[1†]

[1]Department of Physics, Massachusetts Institute of Technology, Cambridge, Massachusetts 02139, USA.
[2]Max Planck POSTECH/Korea Research Initiative, Center for Complex Phase of Materials, Pohang 37673, Republic of Korea.
[3]Department of Physics, Harvard University, Cambridge, Massachusetts 02138, USA.
[4]Department of Physics, University of Toronto, Toronto, Ontario M5S1A7, Canada.
[5]PAL-XFEL, Pohang Accelerator Laboratory, Pohang University of Science and Technology, Pohang 37673, Gyeongbuk, Republic of Korea.
[6]Photon Science Center, Pohang University of Science and Technology, Pohang 37673, Gyeongbuk, Republic of Korea.
[7]Department of Physics, Pohang University of Science and Technology, Pohang 790-784, Republic of Korea.
[8]Canadian Light Source, Saskatoon, SK S7N 2V3, Canada.
[9]Institue for Theoretical Physics, ETH Zürich, Zürich 8093, Switzerland.
[10]These authors contributed equality to the work.
[†]Corresponding author. Email: rcomin@mit.edu

**This PDF file includes:**

    Main Text
    Figures 1 to 5

**Abstract**

Phase fluctuations are a key factor distinguishing nonthermal (ultrafast) and thermal phase transitions. Charge order in cuprates is characterized by short-range coherence while competing with superconductivity, and as such it provides a representative case to study the role of phase fluctuation in coupled order parameter dynamics. In this work, we investigated the intertwined evolution of charge order and superconductivity in cuprate/manganite heterostructures using time-resolved resonant X-ray scattering. The resulting dynamics are analyzed within a space- and time-dependent nonperturbative model capturing both amplitude and phase dynamics. At low fluence, photo-induced suppression of superconductivity results in a nonthermal enhancement of charge order, underscoring the dynamic competition between charge order and superconductivity. With increasing fluence, the slowing down of melting and recovery dynamics is observed, indicating a critical role of phase fluctuations. At high fluence, both charge order and superconductivity remain suppressed for an extended time window due to decoupling between amplitude and phase dynamics and the delayed recovery of phase coherence. Our work underscores the importance of phase fluctuation for understanding the dynamic competition between order parameters in cuprates.

**Significance Statement**

The interplay between competing orders is central to the exotic properties of quantum materials like cuprate unconventional superconductors. While suppressing superconductivity with ultrafast photoexcitation is known to transiently enhance competing charge order, a complete microscopic understanding of this dynamic interplay, particularly the role of phase coherence, has remained elusive. Here, we provide unprecedented insight into these transient phenomena by integrating state-of-the-art time-resolved X-ray scattering with a space- and time-dependent Ginzburg-Landau theory explicitly incorporating *both* amplitude and phase dynamics. Our combined approach reveals the previously underappreciated, yet critical, role of *phase dynamics* in dictating the competition between superconductivity and charge order. This work advances our understanding of competing states in complex materials and highlights the significance of phase coherence ultrafast dynamics.

**Main Text**

A defining trait of strongly correlated materials is the emergence of intertwined orders in their phase diagrams. In the representative case of cuprates, for example, high-temperature superconductivity appears in the phase diagram together with numerous other phases, including antiferromagnetism, spin stripes, charge/spin/pair-density-waves, pseudogap, and strange metal.[1] The relationship between these coexisting orders is complex and multifaceted, yet they are key in shaping the essential physics underlying these materials. For instance, while static charge order in cuprates nominally competes with superconductivity and suppresses the critical temperature $T_c$,[2–4] fluctuating charge order can potentially contribute to electron pairing, stabilizing superconductivity and enhancing $T_c$.[5] However, despite many years of intense research, understanding the rich interplay between intertwined orders in cuprates remains a daunting challenge, one that calls for the new experimental and theoretical approaches.

One effective pathway to elucidate the relationship between intertwined orders is to monitor their relative responses to external perturbations. For example, in cuprates, the evolution of charge order under external stimuli such as pressure,[6,7] magnetic fields,[8,9] uniaxial strain,[10,11] and epitaxial proximity effects[12,13] has been extensively characterized, from which its relationship with neighboring orders has been inferred. In parallel to the study of *(quasi-)static* perturbations, advancements in laser science and X-ray free electron lasers (XFELs) have established ultrafast photoexcitation as an effective tool to *dynamically* perturb the electronic states, with the potential to uncover hidden nonthermal phases not accessible in equilibrium. In the case of cuprates, previous time-resolved resonant X-ray scattering studies (tr-RXS) at XFELs revealed that ultrafast photoexcitation suppresses the superconducting order, leading to a nonthermal enhancement of the competing charge-ordered phase in $YBa_2Cu_3O_{7-d}$.[14,15] Remarkably, this transient charge order state, emerging after the photo-quench of superconductivity, is characterized by not only enhanced intensity but also increased correlation length, providing important information on how the charge-ordered and superconducting domains are competing at the nanoscale prior to photoexcitation. For completeness, we also note that the opposite physics – namely, photoinduced suppression of charge order and counteractive enhancement of superconducting fluctuation – is also observed in $La_{2-x}Ba_xCuO_4$ by time-domain THz spectroscopy, indicating rich pathways to control competing orders by photoexcitation.

In principle, inspecting how charge order and superconductivity affect each other in the time domain can provide rich information on their relationship, the involved degrees of freedom, and their natural timescales.[15–17] For such analysis, however, a detailed microscopic model describing the time evolution of charge and superconducting order parameters needs to be developed and tested against experimental data. In particular, the widely adopted time-dependent Ginzburg-Landau theory (with spatially homogeneous order parameters)[18,19] is fundamentally unsuitable for describing competing orders in cuprates, given the short-ranged, nonuniform nature of charge correlations. Considering the substantial time-dependent variation of the correlation length observed in Ref.[14,15], a dedicated model capable of describing both the order parameter amplitude and correlation length dynamics is necessary to fully elucidate the time-domain competition between charge order and superconductivity in cuprates.

In this work, we investigated the competition of charge order and superconductivity in a cuprate/manganite heterostructure using Tr-RXS and analyzed the resulting dynamics using a space- and time-dependent non-perturbative model. Our model considers not only the order parameter amplitude $\psi(x,t)$ but also the transverse and longitudinal correlation functions, $D_k^{\parallel}(t)$ and $D_k^{\perp}(t)$, as a closed set of dynamical variables, enabling the description of both order parameter and correlation length in the time domain.[20] Fig. 1a illustrates the Tr-RXS experimental geometry[21] and our heterostructure configuration composed of a superconducting $YBa_2Cu_3O_{7-d}$ layer sandwiched between two magnetic $La_{2/3}Ca_{1/3}MnO_3$ layers (LCMO/YBCO/LCMO). Compared to the bulk YBCO samples previously investigated,[14,15] our sandwiched YBCO film offers several advantages for Tr-RXS investigation. First, since the penetration depths of optical and X-ray photons are larger than the sample thickness, our YBCO film is quasi-uniformly illuminated by both the pump and probe pulses, mitigating issues related to pump- and probe-depth mismatch and facilitating comparison with theoretical models. Moreover, the long-range proximity effects from the neighboring LCMO layers enhance the charge order in the YBCO layer (see below), allowing the precise detection of weak transient signals using Tr-RXS. Finally, unlike the bilayer YBCO/LCMO configuration previously studied using static RXS,[22] our trilayer structure symmetrizes the heteroepitaxial strain and maximizes the proximity effect.[23] We used a 1.55 eV near-infrared pulse for photoexcitation and stroboscopically probed the time-dependent diffraction at the charge order Bragg vector using 931.3 eV (Cu-$L_3$ absorption edge) XFEL pulses resonant with electronic states in the $CuO_2$ plane. Additional details on sample fabrication and the Tr-RXS experiment are presented in Supplemental Materials.

Before discussing the Tr-RXS results, we first characterize the charge ordered phase in the YBCO hetero-trilayer using static RXS. Fig. 1b displays the temperature-dependent RXS spectra measured at the Cu-$L_3$ edge. As the temperature decreases, a well-defined charge order peak, characterized by the wave vector $|Q_\parallel| \approx 0.31$ r.l.u. and the correlation length $\xi \approx 35$ Å, develops on top of the smoothly varying fluorescence background. The peak intensity reaches a maximum near the superconducting transition ($T_c \approx 65$ K) and decreases monotonically with temperature in the superconducting state. This suppression of the charge order intensity in the superconducting phase is well-documented in previous studies on bulk YBCO (Ref.[2–4]) and demonstrates the competition between charge order and superconductivity under equilibrium conditions. However, such competition was not observed in previous RXS studies of YBCO/LCMO superlattices[13]. Notably, the charge correlation in our YBCO hetero-trilayer persists up to 220 K, indicating a significant enhancement of the charge order onset temperature $T_{CO}$ compared to the bulk value ($T_{CO} \approx 150$ K; see Fig. 1c). While understanding the exact origin of the enhanced $T_{CO}$ requires further investigation into charge, orbital, and magnetic proximity effects at the interface,[24–27] we note that this is the highest $T_{CO}$ ever observed in the 123 cuprate family and is comparable to the pseudogap transition temperature $T^*$ at this doping (Fig. 1c).[28]

Having characterized the charge correlation in equilibrium, we next discuss the transient response of charge order upon ultrafast photoexcitation (Fig. 2). Fig. 2b,c display the time evolution of the charge order peak amplitude after optical pumping, as measured in the normal state (65 K) and superconducting state (20 K), respectively. For comparison, static charge order peak profiles measured at negative delays are also shown in Fig. 2a. While the time traces measured in the normal state can be simply described by a single exponential component, the evolution of charge order peak in the superconducting state is more complex (Fig. 2c) and encodes key information on the interaction between charge order and superconductivity.

In the normal state (Fig. 2b), fitting the time traces with the standard function $\left(1 + \frac{1}{2} erf\left(\frac{t-t_0}{\sqrt{2}\tau_m}\right)\right) * (Ae^{-(t-t_0)/\tau_r} + B)$ reveals a resolution-limited ultrafast melting within $\tau_m < 100$ fs, followed by exponential recovery within $\tau_r$ of a few picoseconds. This quasi-instantaneous melting of charge order in YBCO is distinct from the phonon-limited melting process in other charge-ordered systems, such as 1$T$-$TaS_2$ ($\tau_m \approx 230$ fs)[29] $CsV_3Sb_5$ ($\tau_m \approx 250$ fs),[30] and $LaTe_3$ ($\tau_m \approx 400$ fs)[31] and provides time-domain evidence of the electronic nature of charge order in cuprates. Meanwhile, the recovery timescale $\tau_r$ increases from 1.2 ps to 2 ps with increasing pump fluences, evidencing a slowing down of the dynamics. This effect, commonly observed in solid-state pump-probe experiments, suggests a crucial role for low-energy fluctuations[18], particularly as the system approaches the critical melting of order.[32] At the highest fluence of 2 mJ/cm², nearly full quenching of charge order is achieved without noticeable recovery within the investigated time window.

In contrast to the normal state behavior, in the superconducting state (Fig. 2c), the ultrafast melting of charge order is followed by a significant recovery and overshoot in the charge order amplitude at time delays of 1-4 ps (see $F = 0.08$ mJ/cm² trace, for example). At the simplest level, one can understand this overshooting as a consequence of the light-induced suppression of the competing superconducting order, transiently creating phase space available for charge order to proliferate. However, it is important to note that the transient enhancement of charge order intensity goes well beyond the maximum peak height observed in static measurements (upper horizontal line in Fig. 2), emphasizing the nonthermal nature of the photo-induced state. The timescale of this transient enhancement (~3 ps for $F$ = 0.08 mJ/cm$^2$) is much slower than the timescales observed in the normal state, suggesting that it is connected to the timescale of the superconducting order. With increasing pump fluences, the overshooting becomes weaker and slower, which can be attributed to the combined effects of the increased $t_r$ and the incomplete intensity recovery $B$ of the charge order.

While the time traces in Fig. 2 track the time evolution of the charge order peak amplitude, a complete understanding of charge order dynamics requires further characterization of its correlation length. For this, we measured the charge order peak profiles at selected time delays, i.e. at $\Delta t$ = –1 ps and 0.2 ps in the normal state (see vertical lines in Fig. 3a) and –1 ps, 0.2 ps, and 2.5 ps in the superconducting state (vertical lines in Fig. 3b). These $\Delta t$ values correspond to the pre-pump peak profile, maximum melting, and maximum overshooting of the peak, respectively. We analyzed the obtained peak profiles using a pseudo-Voigt fit, accounting for both instrumental resolution and intrinsic peak width (Fig. 3c,d) – from this, we derived the correlation length ($\xi$) as well as the integrated intensity ($I$) of the charge order peak as a function of delay time and temperature.

The time evolution of $I$ and $\xi$ are summarized in Fig. 3e,f, respectively. For comparison, we also overlaid their thermal evolutions measured in static RXS as grey lines. Within ~150 fs after photoexcitation, not only the integrated intensity but also the correlation length of the charge order become dramatically suppressed. The suppression in correlation length is particularly notable in the normal state, indicating more than a 30 % reduction in the spatial coherence of the charge ordered state. In the superconducting state, both $I$ and $\xi$ exhibit an overshoot at ~2.5 ps due to the suppression of the competing superconducting phase. Again, the enhancement of the integrated intensity exceeds the maximum intensity in static measurements, highlighting the nonthermal nature of the photoexcited state. Unlike the diffraction intensity, the correlation length at ~2.5 ps remains smaller than the maximum static correlation length, potentially reflecting a slower recovery of the order parameter phase coherence compared to the amplitude.[33,34]

We now discuss the microscopic model to understand the observed time-domain interaction between charge order and superconductivity in cuprates. The charge order dynamics we observed in the YBCO hetero-trilayer are overall comparable to those reported in bulk YBCO crystals.[14,15] In these previous studies, the transient enhancement of charge order has been discussed under several scenarios,[14] including homogeneous order parameter competition, spatial phase separation, and the quenching of superconductivity-induced topological defects. However, the detailed microscopic modeling is yet to be developed. A model that goes beyond conventional time-dependent Ginzburg-Landau theory[18] or displacive excitation of coherent phonons[35] is necessary for describing the physics of competing orders in cuprates, given the short-ranged and spatially varying nature of charge correlation.

Here, we modeled the time-domain interaction between charge order and superconductivity in YBCO by extending time-dependent Ginzburg-Landau theory to incorporate spatial degrees of freedom. In our model, the order parameter $\psi(x,t)$ and dynamical structure factors associated with longitudinal fluctuations $D_k^{\parallel}(t) = \langle \psi_{i,\parallel}(-k,t)\psi_{i,\parallel}(k,t)\rangle_c$ and transverse ones $D_k^{\perp}(t) = \langle \psi_{i,\perp}(-k,t)\psi_{i,\perp}(k,t)\rangle_c$ constitute a closed set of dynamical variables, whose time-evolution is dictated by coupled self-consistent integro-differential equations.[20] We used two order parameters $\psi_{CO}(x,t)$ and $\psi_{SC}(x,t)$ to describe coexisting charge order and superconductivity, and imposed their coupling in the free energy. From this, one can model the transient evolution of competing order parameter amplitudes, $\psi_{CO}(t), \psi_{SC}(t)$, as well as their correlation lengths, $\xi_{CO}(t), \xi_{SC}(t)$, as detailed in Supplemental Materials. We note that a similar model has been applied to study the competing *a*-axis and *c*-axis charge order dynamics in rare earth tritellurides.[36]

Figure 4 summarizes the representative dynamics of $\psi_{CO}, \psi_{SC}, \xi_{CO}$, and $\xi_{SC}$ obtained from our model. Remarkably, our simulation captures all the salient features of the experimental dynamics discussed in Fig. 2 and Fig. 3. These include: 1) the initial ultrafast suppression of both $\psi_{CO}$ and $\xi_{CO}$ upon photoexcitation; 2) the subsequent overshooting of $\psi_{CO}$ in ~2 ps in response to the prolonged suppression of $\psi_{SC}$; 3) the slower and weaker overshooting of $\psi_{CO}$ with increasing pump fluence; and 4) the weaker enhancement of $\xi_{CO}$ compared to $\psi_{CO}$. In particular, our simulation indicates a slower recovery of phase coherence ($\xi_{CO}$) compared to the order parameter amplitude ($\psi_{CO}$), highlighted with filled circles in Fig. 4a,c. This separation between amplitude and phase timescales, referred to as 'mode decoupling' in Ref.[20], reflects the overpopulated transverse Goldstone mode $D_k^\perp$ and plays a central role in determining the latter time dynamics according to self-similar evolution. In our case, this mode decoupling and slower recovery of phase coherence accounts for the absence of overshooting in $\psi_{CO}$ at intermediate fluence (F3 in Fig. 4), despite the long-lasting suppression of $\psi_{SC}$, as the $\xi_{CO}$ also remains suppressed for longer timescales (see open circles in Fig. 4a-c). Overall, our simulation underscores the importance of phase dynamics in capturing the detailed time-domain evolution of charge order in superconducting cuprates. The close correspondence obtained between our model and experimental results provides the groundwork for understanding competing order parameter dynamics in cuprates and relevant materials.

Finally, we investigate the charge order dynamics in YBCO hetero-trilayers as a function of YBCO thickness *d*. First, from the static RXS measurements presented in Fig. 5a-c, we observed that the charge order wave vector $Q_{CO}$ shifts toward higher value with decreasing *d*, from $Q_{CO}$ = 0.305±0.005 r.l.u. for *d* = 50 nm to 0.323±0.01 r.l.u. for *d* = 10 nm. According to the established relationship between $Q_{CO}$ and *p* in YBCO,[4,28] this indicates a decrease of hole doping *p* with decreasing *d*, consistent with interface-induced change of hole concentration.[37] In this context, by varying the thickness *d*, we can effectively explore the charge order dynamics across the doping axis of the YBCO phase diagram (see Fig. 1c). Now comparing their dynamics (Fig. 5d-f), we find that the overshooting of the charge order amplitude is strongest at *d* = 50 nm and decreases with decreasing *d* and hence decreasing *p*. Since the strength of superconductivity is maximal in the *d* = 50 nm film, this suggests that the largest portion of charge order, suppressed in equilibrium due to competition with superconductivity, is transiently recovered at this doping level. Despite such doping dependence, the overshooting of the charge order peak height is nonetheless observed in all our investigated hetero-trilayers, confirming that the time-domain interplay between charge order and superconductivity is prevalent across the phase diagram.

In conclusion, we present a systematic tr-RXS study of charge order dynamics in cuprate/manganite heterostructures as a function of temperature, fluence, and thickness (doping). The obtained amplitude and correlation length dynamics are qualitatively reproduced using space- and time-resolved Ginzburg-Landau theory, highlighting the importance of phase coherence and order parameter fluctuations in describing the transient evolution of charge order and superconductivity. In general, how phase coherence is destroyed and re-established after a photo-quench of the order parameter is an integral topic of nonequilibrium dynamics, as such process in ultrafast setting can be fundamentally different from the equilibrium condition, involving, for example, the overpopulation of topological defects based on the Kibble-Zurek mechanism. We propose that our model, capable of describing both amplitude and correlation length dynamics, is applicable to a wide-range of time-resolved studies involving the ultrafast quench and recovery of one or more short-ranged order parameters.

**Materials and Methods**

**Sample growth and characterizations** The LCMO/YBCO/LCMO trilayers used in this study were grown on SrTiO$_3$ substrates using pulsed laser-ablated deposition (PLD). PLD growth was done using a KrF excimer laser operating at a 2J/cm$^2$ fluence. The growth was done at a substrate temperature of 800 °C and 200 mTorr O$_2$ chamber pressure. After deposition, the films were slowly cooled down to 300 °C over 45 minutes in 150 Torr of O$_2$. This underdoping was done to deliberately under-oxygenate the YBCO layer to achieve a doping level that maximized the extent of the charge order in the 50 nm trilayer. LCMO layer thickness was kept at 10 nm, while the YBCO layer thickness was varied from 10 nm to 50 nm. Layer thicknesses were set using material-specific growth rates. These growth rates were determined using atomic force microscopy (AFM) on chemically etched single-layer films and corroborated using x-ray reflectometry (XRR) measurements. The as-grown films were characterized by transport measurements using an AC resistance bridge in the Kelvin configuration. Superconducting transitions were observed at Tc = 70 K, 63 K, and 43 K for d = 50 nm, 30 nm, and 10 nm, respectively. This trend in Tc indicates a decrease in hole concentration with decreasing *d*, consistent with the calibration using $Q_{CO}$ in the Main text. The origin of the hole concentration change in our LCMO/YBCO/LCMO trilayers is investigated in detail in Ref. 23, revealing deoxygenation effects due to heteroepitaxial strain and lattice mismatch as the primary factors.

**Static RXS measurements at synchrotron** RXS experiments were performed at the REIXS beamline of the Canadian Light Source. The samples were cooled using a closed-cycle LHe cryostat in ultrahigh vacuum chamber with base pressure better than 10$^{-9}$ Torr. The films were oriented *in situ* using YBCO Bragg reflections at non-resonant conditions. All RXS measurements were conducted using 931.3 eV photons resonant to the Cu *L*$_3$ absorption edge. The photon polarization was fixed to the out-of-scattering plane direction (s polarization) to maximize the sensitivity to charge scattering. The in-plane momentum was scanned by rocking the sample angle at fixed detector position. For the temperature dependence in Fig. 1b, we waited at least 30 mins after each temperature change to ensure thermal equilibrium.

**Time-resolved RXS measurements at XFEL** Tr-RXS experiments were conducted at the RSXS beamline of the PAL-XFEL. The samples were cooled using an open-cycle LHe cryostat within an ultrahigh vacuum chamber with a base pressure at the level of 10$^{-9}$ Torr. The photon polarization was fixed parallel to the scattering plane (*p* polarization), but still yielded sufficient charge order reflection to observe transient changes. We used 1.55 eV NIR pulses produced by a Ti:Sapphire laser and 931.3 eV XFEL pulses tuned to the Cu *L*$_3$-edge as the pump and probe pulses, respectively. The repetition rate of the pump pulse was set to half (30 Hz) of that of the probe pulse (60 Hz), so that half of the XFEL pulses recorded the photoexcited state while the remaining half recorded the unperturbed state as a reference (Supplementary Fig. S1). The optical pump pulses were injected into the scattering chamber nearly parallel to the XFEL pulses. The footprint of the pump beam (~500 · 500 mm$^2$) was set to be much larger than the probe beamspot (~100 · 200 mm$^2$) to ensure quasi-uniform photoexcitation across the entire probe volume.

**Space- and Time-dependent Ginzburg-Landau simulation** The space- and time-dependent Ginzburg-Landau simulation used in this work is an extension of the model developed in Ref.[20], incorporating two competing order parameters. Unlike the widely used time-dependent Ginzburg-Landau model with spatially homogeneous order parameters, $\psi(t)$, our model considers space-dependent order parameters, $\psi(x,t)$, to account for the short-ranged nature of charge correlations in cuprates. The dynamics of these space-dependent order parameters are parametrized by two observables, the time evolution of the order parameter strength, $\psi(t)$, and the time evolution of the respective dynamical structure factors $D_k^{\parallel}(t)$ and $D_k^{\perp}(t)$, from which one can extract the dynamics of the correlation length $\xi(t)$.

The Ginzburg-Landau free energy functional describing two competing order parameters (charge order and superconductivity in our case) has the form:

$$F[\psi_{CO},\psi_{SC}] = F_{CO}[\psi_{CO}] + F_{SC}[\psi_{SC}] + F_{int}[\psi_{CO},\psi_{SC}],$$

where $F_i[\psi_i]$ is the usual Mexican-hat potential:

$$F_i[\psi_i] = \int d^d x \left[\frac{r_i}{2}\psi_i^2 + u_i\psi_i^4 + \frac{K_i}{2}\nabla\psi_i^2\right],$$

and $F_{int}[\psi_{CO}, \psi_{SC}]$ describes the interaction between two orders:

$$F_{int}[\psi_{CO}, \psi_{SC}] = 2\eta \int d^d x \psi_{CO}^2 \psi_{SC}^2.$$

Practically, $r_i$ and $u_i$ determine the shape of the Mexican-hat potential, $K_i$ determines the order parameter stiffness, and $\eta$ governs the strength of the competition, with $\eta > 0$.

Meanwhile, both charge order and superconductivity can be described by the complex two component order parameters (real and imaginary parts of $\psi_i$):

$$\psi_i = [\psi_{i,1}, \psi_{i,2}].$$

Without loss of generality, we assumed that spontaneous symmetry breaking occurs along the first component, $\psi_{i,1}$, i.e.

$$\psi_{i,1}(t) = \langle \psi_{i,1}(t) \rangle.$$

Then fluctuations occurring parallel and perpendicular to the spontaneous symmetry breaking represent longitudinal (Higgs) and transverse (Goldstone) excitations, respectively:

$$D^{\parallel}_{k,i}(t) = \langle \psi_{i,\parallel}(-k,t)\psi_{i,\parallel}(k,t) \rangle_c = \langle \psi_{i,1}(-k,t)\psi_{i,1}(k,t) \rangle_c$$

$$D^{\perp}_{k,i}(t) = \langle \psi_{i,\perp}(-k,t)\psi_{i,\perp}(k,t) \rangle_c = \langle \psi_{i,2}(-k,t)\psi_{i,2}(k,t) \rangle_c.$$

These three parameters – the order parameter amplitude $\psi_{i,1}(t)$, longitudinal dynamical structure factor $D^{\parallel}_{k,i}(t)$, and transverse one $D^{\perp}_{k,i}(t)$ – form a closed set of dynamical variables that completely describes the time evolution of the system. The equations of motion for these parameters are (model A in notation of Ref.[38])

$$\frac{d\psi_i(t)}{dt} = -\Gamma_i r_{eff,i} \psi_i,$$

$$\frac{dD^{\perp}_{k,i}(t)}{dt} = 2T\Gamma_i - 2\Gamma_i(K_i k^2 + r_{eff,i}) D^{\perp}_{k,i},$$

$$\frac{dD^{\parallel}_{k,i}(t)}{dt} = 2T\Gamma_i - 2\Gamma_i(K_i k^2 + r_{eff,i} + 8u_i\psi_i^2) D^{\parallel}_{k,i}.$$

where $r_{eff,i}$ is a self-consistent "mass term", defined as

$$r_{eff,co}(t) = r_{CO} + 4u_{CO}\left[\psi_{CO}^2 + \int \frac{d^3k}{(2\pi)^3}(D_{k,CO}^{\parallel} + D_{k,CO}^{\perp})\right] + 4\eta\left[\psi_{SC}^2 + \int \frac{d^3k}{(2\pi)^3}(D_{k,SC}^{\parallel} + D_{k,SC}^{\perp})\right],$$

and vice versa for $r_{eff,SC}$.

One can understand the meaning of $r_{eff,i}$ as follows. In the simplest case, where there exists only one spatially homogeneous order parameter, the pump-induced suppression and recovery of the order parameter can be simply described by

$$\frac{d\psi_{CO}(t)}{dt} = -\Gamma r\psi_{CO},$$

which results in single exponential recovery dynamics. In contrast, in the case of spatially inhomogeneous order parameter, the recovery of $\psi_{CO}$ depends not only on the magnitude of $\psi_{CO}$ at a given $t$, but also on the fluctuations, $D_{k,CO}^{\parallel}$ and $D_{k,CO}^{\perp}$. The order parameter and fluctuation dynamics thus exhibit coupled evolution; this coupling is described in our model by replacing the bare $r$ with $r_{eff}$, which contains contributions from fluctuation as displayed in the second term in the expression of $r_{eff}$. Note that $r_{eff}$ itself exhibits complex time dependence and directly determines the dynamics of $\psi_i$ according to the equation $\frac{d\psi_i(t)}{dt} = -\Gamma_i r_{eff,i}\psi_i$. In addition to the fluctuation effects, coupling to superconducting order (with coupling strength $\eta$) also influences the dynamics of charge order. This coupling is described by the last term in $r_{eff}$.

We note that, compared to the standard time-dependent Ginzburg-Landau theory with a spatially homogeneous order parameter, our model introduces only two additional parameters, $K_{CDW}$ and $K_{SC}$, describing how the free-energy is affected by the spatial variance of the order parameters $\nabla\psi_{CDW}$ and $\nabla\psi_{SC}$. These additional parameters are determined to fit independent experimental inputs, namely the time-dependence of the coherence length. Thus, despite the larger number of parameters compared to the spatially homogeneous time-dependent Ginzburg-Landau model, our theory remains well-constrained.

The above coupled equations of motion can be solved numerically using standard ODE solvers. Depending on the parameters used, the free energy functional $F[\psi_{CO}, \psi_{SC}]$ describes either a bicritical regime ($u_{CO}u_{SC} > \eta^2$), where only charge order develops in equilibrium, or a tetracritical regime ($u_{CO}u_{SC} < \eta^2$), where both charge order and superconductivity coexist.[39] The competing dynamics of charge order and superconductivity obtained in the tetracritical regime are displayed in Fig. 4. For completeness, we also describe the dynamics of charge order in the bicritical regime in Supplementary Fig. S2.

**Acknowledgments**

We acknowledge Canadian Light Source and Pohang Accelerator Laboratory for provision of synchrotron and XFEL beamtime. This work was primarily supported by the Department of Energy, Office of Science, Office of Basic Energy Sciences, under Award Number DE-SC0019126. The tr-RSXS experiments were performed at the RSXS endstation (proposal number: 2021-1st-SSS-013) of the PAL-XFEL, funded by the Korea government (MSIT). Work at the University of Toronto was supported by the Natural Sciences and Engineering Research Council (NSERC) through the Discovery Grant and the Alliance International Catalyst Quantum Grant, and by the Canada Foundation for Innovation (CFI). M.K., B.L., J.-H.P., and H.J. acknowledges the support by the National Research Foundation of Korea (NRF) funded by the Ministry of Science and ICT(No. RS-2022-NR068223). Research performed in Canadian Light Source is funded by the Canada Foundation Innovation, the Natural Sciences and Engineering Research Council of Canada, the University of Saskatchewan, the Government of Saskatchewan, Western Economic Diversification Canada, the National Research Council Canada, and the Canada Institutes of Health Research.

**Figures and Tables**

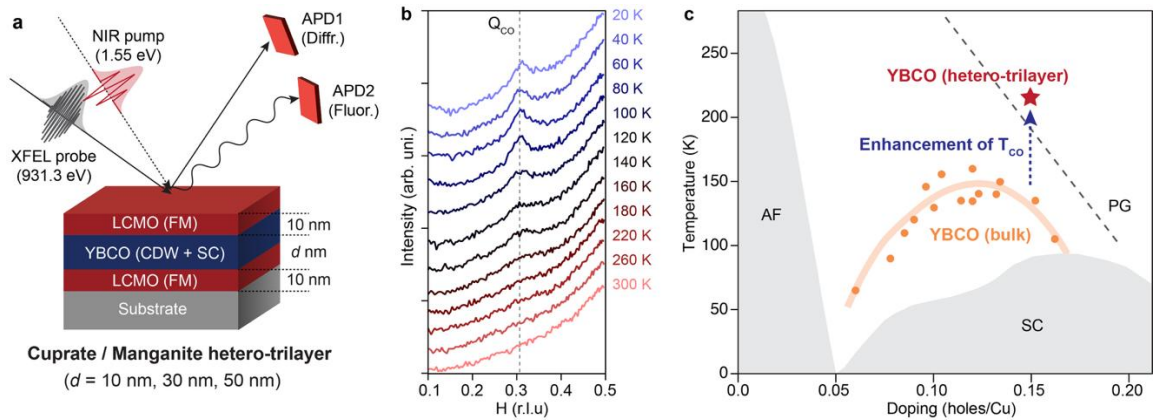

**Figure 1. Enhancement of charge order in cuprate/manganite heterotrilayers. a,** Schematics of time-resolved RXS experimental geometry. Near-colinear 1.55 eV NIR pulse and 931.3 eV XFEL pulse were used as pump and probe beams, respectively. Diffracted X-ray photons and fluorescence signal were recorded using two avalanche photodiodes (APD), placed on and off the scattering plane, respectively. The bottom schematic shows the LCMO/YBCO/LCMO heterotrilayer sample configuration. **b,** Temperature-dependent static RXS spectra detecting the charge order reflection in LCMO/YBCO/LCMO heterotrilayer. **c,** The phase diagram of bulk YBCO, showing the superconductivity (SC), antiferromagnetism (AF), pseudogap (PG), and charge order phases (orange symbols). The overlaid red star indicates the onset of charge order phase in our LCMO/YBCO/LCMO heterotrilayer, highlighting the significant enhancement of $T_{CO}$ compared to the bulk value.

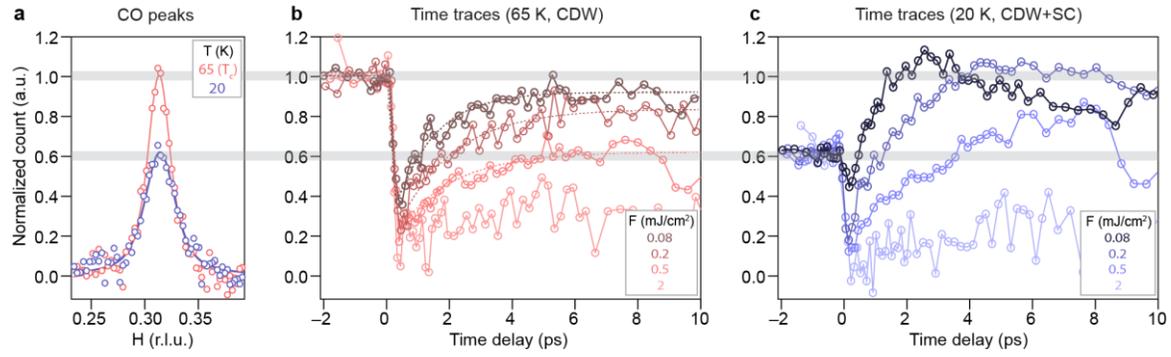

**Figure 2. Transient evolution of the charge order peak amplitude after photoexcitation. a,** Static charge order peak profiles of $d$ = 30 nm film measured at XFEL at negative time delay. The horizontal grey lines across the panels mark the peak amplitudes at 65 K ($T_c$) and 20 K. All data in this figure are normalized with respect to the peak amplitude at 65 K to facilitate comparison. The overlaid solid lines are fits to pseudo-Voigt distribution. **b,c,** Time evolution of the charge order peak amplitude measured at the normal state (65 K) and the superconducting state (20 K), respectively. The overlaid dashed lines in b are fit to the single exponential function discussed in the main text.

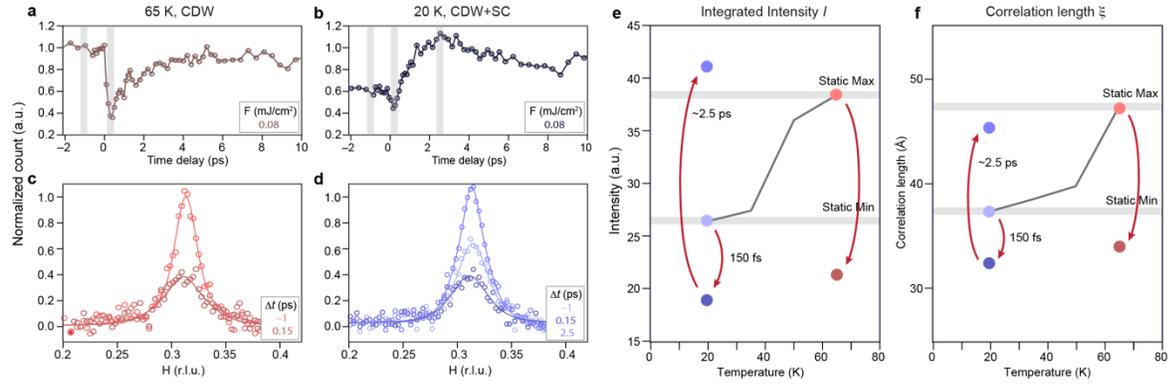

**Figure 3. Transient evolution of the charge order peak coherence after photoexcitation. a,b,** Time traces of the charge order peak amplitudes at 65 K and 20 K, reproduced from Fig. 2. The vertical grey lines indicate characteristic time delays corresponding to the negative delay, maximum peak suppression, and maximum peak overshooting, respectively. **c,d,** Charge order peak profiles measured at the time delays marked in a,b. The overlaid solid lines are fits to a pseudo-Voight distribution, from which the integrated peak intensities and correlation lengths were obtained. **e,f,** Summary of the thermal and nonthermal evolution of the charge order peak intensity and correlation length, respectively. The horizontal grey lines mark the maximum and minimum values in the static limit.

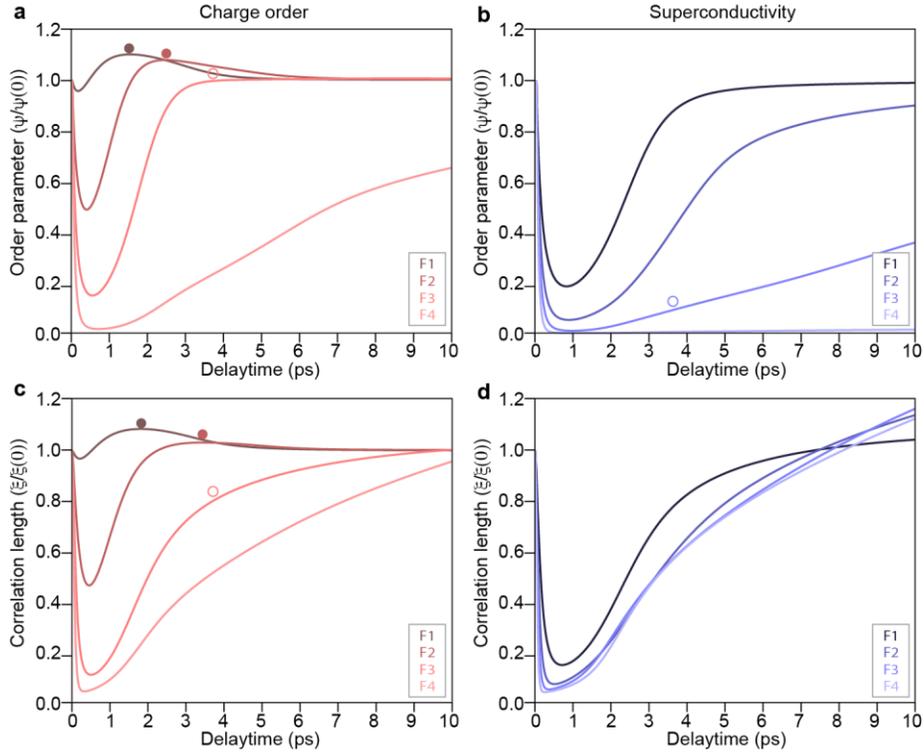

**Figure 4. Simulation of order parameter and correlation length dynamics using space- and time-dependent Ginzburg-Landau model. a,c,** Order parameter and correlation length dynamics of charge order phase at four reference pump fluences. **b,d,** Corresponding parameter and correlation length dynamics of superconducting phase. Filled circles in a,c mark the time delay corresponding to the maximum overshooting: the separation of timescales between order parameter and correlation length is apparent from this comparison. Open circles in a,b,c compare the recovery of $\psi_{co}$, $\psi_{sc}$, and $\xi_{co}$ at identical delaytime. The recovery of $\xi_{co}$ is clearly slower than the recovery of $\psi_{co}$, limiting the overshooting at the intermediate fluence despite the prolonged suppression of $\psi_{sc}$.

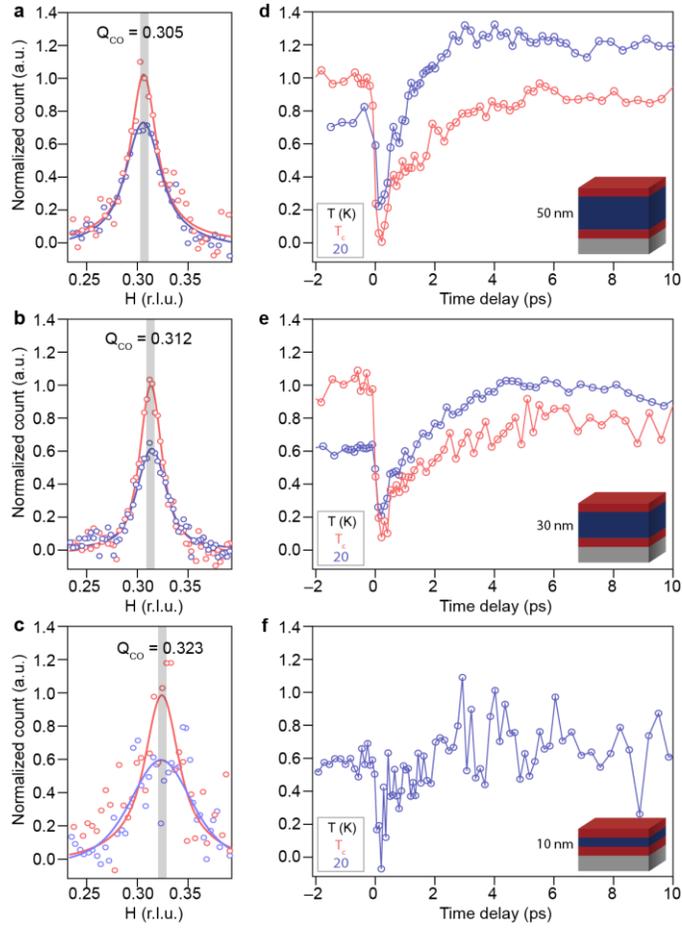

**Figure 5. Thickness dependence of charge order dynamics. a-c,** Charge order peak profiles of $d$ = 50, 30, and 10 nm films, respectively, measured at negative time delay. The charge order wave vector $Q_{CO}$ shifts toward higher values with decreasing $d$, indicating a reduction of hole concentration in thinner films. **d-f,** Time traces of the charge order amplitude in $d$ = 50, 30, and 10 nm films, measured at $T_c$ and at the base temperature. The pump fluence were fixed to 0.2 mJ/cm$^2$ across all measurements. The insets in d-f show schematics of the heterotrilayer configuration.

**Supporting Information for**

**Role of phase fluctuation in dynamic competition between charge order and superconductivity in cuprates**

Mingu Kang[1,2,10], Pavel E. Dolgirev[3,10], Chao C. Zhang[4], Hoyoung Jang[5,6], Byungjune Lee[2,7], Minseok Kim[5], Sang-Youn Park[5], Ronny Sutarto[8], Eugene Demler[9], Jae-Hoon Park[2,6,7], John Y. T. Wei[4] & Riccardo Comin[1†]

[1]Department of Physics, Massachusetts Institute of Technology, Cambridge, Massachusetts 02139, USA.
[2]Max Planck POSTECH/Korea Research Initiative, Center for Complex Phase of Materials, Pohang 37673, Republic of Korea.
[3]Department of Physics, Harvard University, Cambridge, Massachusetts 02138, USA.
[4]Department of Physics, University of Toronto, Toronto, Ontario M5S1A7, Canada.
[5]PAL-XFEL, Pohang Accelerator Laboratory, Pohang University of Science and Technology, Pohang 37673, Gyeongbuk, Republic of Korea.
[6]Photon Science Center, Pohang University of Science and Technology, Pohang 37673, Gyeongbuk, Republic of Korea.
[7]Department of Physics, Pohang University of Science and Technology, Pohang 790-784, Republic of Korea.
[8]Canadian Light Source, Saskatoon, SK S7N 2V3, Canada.
[9]Institue for Theoretical Physics, ETH Zürich, Zürich 8093, Switzerland.
[10]These authors contributed equality to the work.
[†]Corresponding author. Email: rcomin@mit.edu

**This PDF file includes:**

    Figures S1 and S2

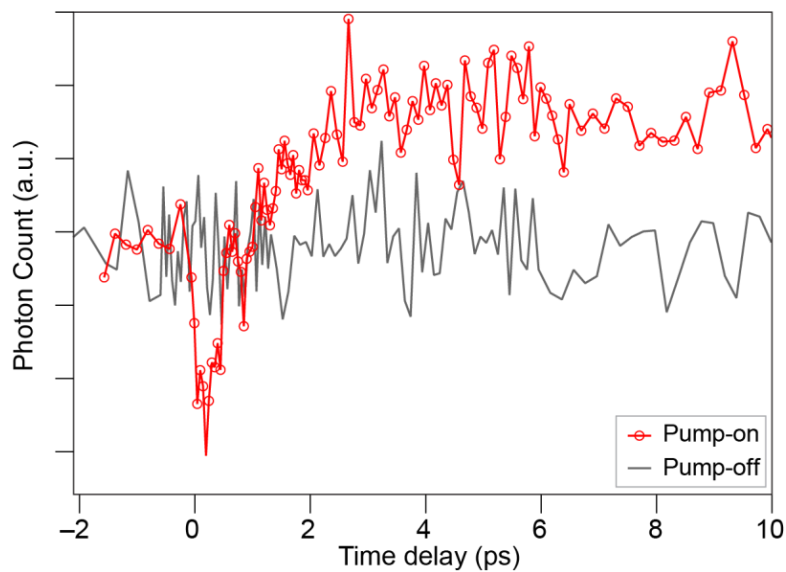

**Figure S1. Comparison between pump-on and pump-off signals in tr-RXS experiments**. As described in the Methods section, our Tr-RXS measurements were conducted in a configuration that the repetition rate of the pump pulse was set to half (30 Hz) of that of the probe pulse (60 Hz). This means that one of two XFEL pulses recorded the photoexcited state, while the other pulse recorded the unperturbed state as a reference. The collected data were then separated into either pump-on or pump-off signals on a shot-to-shot basis. This measurement configuration allows us to diagonize the contribution from external factors – such as static heating, beam damage, and drift in sample positions – by monitoring the pump-off signals. In this figure, we display representative pump-on and pump-off signals of the charge order reflection obtained from the $d$ = 50 nm hetero-trilayer film at 20 K. The pump-off signal exhibits negligible variation across the acquisition time (~ 30 min), allowing us to rule out influences from extrinsic factors.

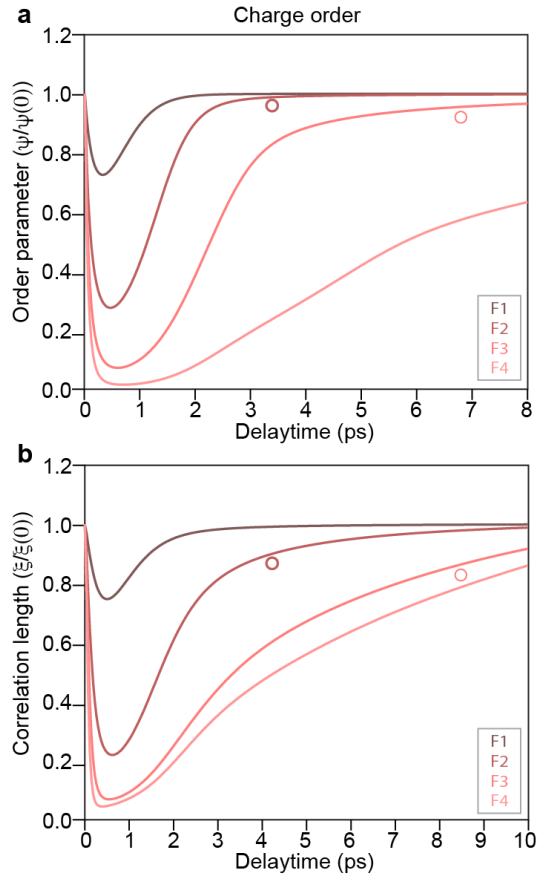

**Figure S2. Charge order dynamics in a bicritical regime**. In this regime ( $u_{co}u_{sc} > \eta^2$ ), the superconducting order is absent in equilibrium, and our model describes the dynamics of pure charge ordered state. Notably, the simulation captures the slowing down of both melting and recovery time scales of $\psi_{co}$ with increasing pump fluences; the former and the latter are often referred to as 'dynamical slowing down' and 'critical slowing down' in the literature, respectively. In addition, the separation of time scales in order parameter amplitude (**a**) and phase coherence (**b**) is reproduced, as discussed in the Main text. To illustrate this, the open circles mark the time delay where the order parameter amplitude is almost fully recovered (**a**), while at the same time delay, the phase coherence is still recovering (**b**).